\documentclass[12pt, draftclsnofoot, onecolumn]{IEEEtran}
\IEEEoverridecommandlockouts
\usepackage{cite}
\usepackage{amsmath,amssymb,amsfonts}
\usepackage{bm}
\usepackage{algorithmic}
\usepackage[linesnumbered,ruled,vlined]{algorithm2e}
\usepackage{graphicx}
\usepackage{textcomp}
\usepackage{xcolor}
\usepackage{capt-of}  
\usepackage{cuted}    
\usepackage{multirow}
\usepackage{lipsum}
\usepackage{hyperref}
\def\BibTeX{{\rm B\kern-.05em{\sc i\kern-.025em b}\kern-.08em
    T\kern-.1667em\lower.7ex\hbox{E}\kern-.125emX}}
\DeclareMathOperator*{\argmax}{arg\,max}
\begin{document}

\title{Graph Neural Networks-Based User Pairing in Wireless Communication Systems\\}

\author{\IEEEauthorblockN{Sharan Mourya,} \thanks{The authors are with the Department of Electrical Engineering, Indian Institute of Technology Hyderabad, India.
(email: sharan.mourya@5g.iith.ac.in, \quad pavan@wisig.com, \quad asaidhiraj@ee.iith.ac.in, \quad kkuchi@ee.iith.ac.in).}
\and
\IEEEauthorblockN{Pavan Reddy,}
\and
\IEEEauthorblockN{SaiDhiraj Amuru,}
\and
\IEEEauthorblockN{Kiran Kumar Kuchi}
}


\maketitle

\begin{abstract}
Recently, deep neural networks have emerged as a solution to solve NP-hard wireless resource allocation problems in real-time. However, multi-layer perceptron (MLP) and convolutional neural network (CNN) structures, which are inherited from image processing tasks, are not optimized for wireless network problems. As network size increases, these methods get harder to train and generalize. User pairing is one such essential NP-hard optimization problem in wireless communication systems that entails selecting users to be scheduled together while minimizing interference and maximizing throughput. In this paper, we propose an unsupervised graph neural network (GNN) approach to efficiently solve the user pairing problem. Our proposed method utilizes the Erdős goes neural pipeline to significantly outperform other scheduling methods such as $k$-means and semi-orthogonal user scheduling (SUS). At 20 dB SNR, our proposed approach achieves a $49\%$ better sum rate than $k$-means and a staggering $95\%$ better sum rate than SUS while consuming minimal time and resources. The scalability of the proposed method is also explored as our model can handle dynamic changes in network size without experiencing a substantial decrease in performance. Moreover, our model can accomplish this without being explicitly trained for larger or smaller networks facilitating a dynamic functionality that cannot be achieved using CNNs or MLPs.
\end{abstract}

\begin{IEEEkeywords}
Graph neural networks,   maximum-clique,  semi-orthogonal user scheduling, and user pairing.
\end{IEEEkeywords}

\section{Introduction}

User pairing is an important problem in wireless communication systems that involves selecting users to be scheduled together in a way that minimizes interference and maximizes the sum rate \cite{sus}. Multi-user multiple-input multiple-output (MU-MIMO) technology is designed to provide separate streams of data to multiple users. This means that each user receives only a portion of the transmitted data. To minimize interference between users, it is important to create a group of users that can be scheduled together. While many scheduling algorithms focus on creating subsets of users with orthogonal or semi-orthogonal MIMO channels \cite{sus}, the optimal user pairing solution would require an exhaustive search of all possible combinations of users, which is a challenging problem as it is NP-hard \cite{sus}. In recent times, deep learning techniques have been acknowledged as a promising means of solving such complex problems in wireless communication systems \cite{ocean,mimo,stnet,power,beam,resource}. Channel state information (CSI) prediction task is achieved by using a combination of convolutional neural networks (CNNs) and a long short term with memory (LSTM) network \cite{ocean}. Similarly, a novel CSI feedback network was designed \cite{stnet} in using transformers and a spatially separable attention mechanism. In \cite{mimo}, the authors demonstrated a single link MIMO communication system using unsupervised learning on autoencoders framework. Optimal power control \cite{power}, beamforming \cite{beam}, and resource allocation \cite{resource} have also been solved using deep learning techniques achieving state-of-the-art performances. Consequently, deep learning is regarded as the quintessential technology to drive sixth-generation (6G) wireless communications. 

The prominence of deep learning in varied disciplines is due to factors like the availability of large training datasets, rapidly developing computational resources (such as GPUs), and deep learning’s efficacy in extracting latent representations from data types like text, and images, which are considered to be Euclidean. For instance, images occupy a rectangular subspace in Euclidean space, and CNNs can exploit the spatial invariance and locality of the features embedded in images to extract meaningful latent representations. Nevertheless, there have been numerous applications where data cannot be represented in Euclidean space and the usage of graphs to represent such data becomes essential. Wireless networks also belong to the category of applications where data cannot be efficiently represented in Euclidean space. Social networks, drug discovery, and citation networks are a few other examples \cite{survey}. In recent studies \cite{survey}, graph neural networks (GNNs) have been utilized in analyzing complex real-world networks. GNNs are a novel type of data-driven model that provides a useful framework for the representation learning of graphs. They can also accurately learn and replicate intricate behaviors in graphs\cite{base}. These desirable properties of GNNs make them a promising candidate to solve complex problems in wireless communication systems. In addition, the properties of GNNs are well-suited to meet the needs of wireless communication systems. One key advantage is that GNNs can efficiently utilize graph data for various learning purposes. Additionally, they are capable of accommodating input graphs of different dimensions, which is particularly valuable given the constantly changing nature of wireless communication systems.

While deep learning techniques have successfully addressed many difficult problems in wireless communication, early implementation of neural network models using CNNs and multi-layer perceptrons (MLPs) often struggle to represent wireless networks and scale effectively for large communication networks, owing to the limited generalization and interpretability of these models \cite{gnn2}\cite{gnn1}. GNNs employ message passing between the nodes, allowing them to learn the underlying distribution of graph-structured data at the node/edge/graph level, making GNNs a popular analytical tool in diverse fields. Consequently, GNNs have been introduced to address several optimization problems in wireless communication systems \cite{overview}. Y. Shen et al \cite{gnnw1} used GNNs to efficiently solve wireless power control for a k-user interference channel. A similar approach was taken in \cite{gnnw2} to achieve adaptive power control. GNNs also facilitate decentralized inferring of data as studied in \cite{dec1} \cite{dec2}. Decentralized processing may provide a more efficient implementation of various communication algorithms. One such procedure for power control was studied in \cite{dec1}. In \cite{gnnw3}, the authors demonstrated scalable power control and beamforming techniques using GNNs which are also considered to be scalable as they can effectively handle large-scale graph data even when only trained on smaller graphs. The primary reason for this is that GNNs share the same set of parameters across all nodes, allowing them to extend the trained parameters directly to an unseen node. We will utilize this representational power of GNNs to perform dynamic user pairing where new users can also be scheduled by the model without being explicitly trained. 

In this paper, we propose to use GNNs to solve user pairing problem in real time. Previously, there are several deep-learning approaches proposed to address this problem. $k$-means clustering \cite{prev1} is one popular method that involves partitioning users into groups based on channel characteristics and selecting users from different groups to maximize the sum rate. However, $k$-means clustering does not take into account the correlation between users in different groups, which can lead to sub-optimal scheduling. In other words, interference across users from different groups can decrease MIMO efficiency. In addition, $k$-means clustering may not be suitable for larger networks with many users. Authors in \cite{prev2} achieved user grouping by projecting the statistical CSI of all users into a two-dimensional grid and utilizing the objective detection model you look only once (YOLO) to divide the users into different groups. Although the method is fast, it lacks the desired accuracy and is also not generalizable to more complex 3D channel models. To address these issues, we propose an unsupervised graph neural networks-based approach to scalable user pairing which is fast, accurate, and suitable for large-scale networks. Our contributions can be summarized as follows:

\begin{enumerate}
    \item We introduced a novel distance metric for graph-structured user data called \textit{sum rate distance} to effectively solve user pairing. This distance metric is designed in a way to capture the correlation between the users and their compatibility to be grouped together. 
    \item Using \textit{sum rate distance}, we formulated user pairing as a $k$-clique optimization problem which along with beamforming solves the user pairing problem in polynomial time.
    \item To the best of our knowledge, we are the first to adopt the popular \textit{"Erdős goes neural"} pipeline to solve user pairing.   
    \item We modified the \textit{"Erdős goes neural"} pipeline, which was originally designed for maximum-clique optimization to solve $k$-clique optimization and in turn user pairing. 
\end{enumerate}

The organization of this paper is as follows: In Section II, we introduce the channel model used in this paper. In Section III, we briefly summarize the terms and definitions required to understand the paper. We then formulate the user pairing problem using graph theory in Section IV followed by the framework used to solve this problem using GNNs in Section V. We provide the numerical results of various simulations in Section VI while concluding this paper in Section VII.

\section{Channel Model}
\begin{figure}[h]
\centering
\includegraphics[width=0.5\linewidth]{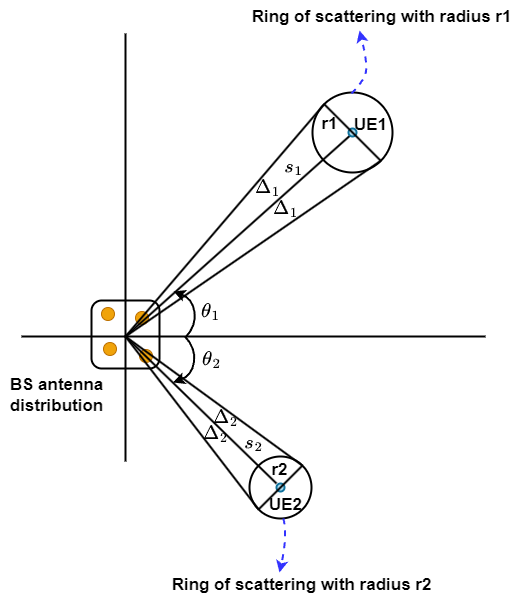}
\captionof{figure}{One-ring channel model with two users illustrating the rings of scattering, azimuthal and radial components.} 
\label{channel}
\end{figure}
To provide a more qualitative discussion on user pairing, we first define the channel model that is used in this paper. Consider a cell with $K$ users in a frequency division duplex (FDD) system with $M$ transmit antennas and each user with one antenna. We then assume the channel to be narrowband in nature and consider the one-ring model \cite{jsdm} as shown in Fig.{~\ref{channel}}, where the $i$-th user equipment (UE) is located at a radial distance of $s_{i}$, and at an azimuthal angle of $\theta_{i}$ that is surrounded by a ring of scatterers of radius $r_{i}$ such that $\Delta_{i} \approx arctan(r_{i}/s_{i})$. The correlation matrix $\bm{R}_{k}$ of the $k$-th user is given by
\begin{equation}
    [\bm{R}_{k}]_{m,p} = \frac{1}{2\Delta_{k}}\int_{-\Delta_{k}}^{\Delta_{k}}e^{j\bm{k}_{k}^T(\alpha_{k}+\theta_{k})(\bm{u}_{m}-\bm{u}_{p})} d\alpha_{k} \nonumber,
\end{equation}
where $[\bm{R}_{k}]_{m,p}$ denotes the correlation between the $m$-th and $p$-th transmit antennas located at the base station (BS) with positional vectors $\bm{u}_{m}, \bm{u}_{p} \in \mathbb{R}^2$. Also, $\bm{k}_{k}(\alpha_{k}) = -\frac{2\pi}{\lambda} (\cos(\alpha_{k}), \sin(\alpha_{k}))^T$ is the wave propagation vector with an angle of arrival of $\alpha_{k}$ and wavelength $\lambda$. Now, let's aggregate $K$ users channel vectors by columns $\bm{H} = [\bm{h}_{1}^T, \hdots, \bm{h}_{K}^T]^T$ where $\bm{H}$ denote the $M \times K$ channel matrix. Thus, the received signal $\bm{y}$ is formulated as

\begin{equation}
    \bm{y} = \bm{H}\bm{x} + \bm{z} \nonumber,
\end{equation}
where $\bm{x}$ is the transmitted symbol, and $\bm{z} \sim \mathbb{C}\mathcal{N}(0,I)$ denotes the complex additive white Gaussian noise (AWGN). Performing the Karhunen Loeve (KL) decomposition \cite{jsdm} on $\bm{R}_{k}$ gives $\bm{R}_{k} = \bm{U}_{k}\bm{\Lambda_{k}}\bm{U}_{k}^{*}$,
where 
$\bm{\Lambda}_{k}$ is an $r \times r$ diagonal matrix whose elements are the non-zero eigenvalues of $\bm{R}_{k}$, and $\bm{U}_{k} \in \mathbb{C}^{M\times r}$ is the tall unitary matrix of the eigenvectors of $\bm{R}_{k}$ corresponding to the non-zero eigenvalues and $r$ is the rank of $\bm{R}_{k}$.

\section{Background}
In this section, we introduce the terms and definitions necessary to understand this paper. We also motivate the usage of GNNs for solving user pairing.
\subsection{User Pairing}
Let $\mathcal{K}$ be the set of $K$ users in the system, and $\mathbb{K}_{k}$ be an arbitrary subset of $\mathcal{K}$ with size $k$. We formulate the user pairing problem for finding the optimal subset $\mathbb{K}^*_{k}$ which achieves the maximum capacity for a given $k$ as follows
\begin{equation}
    \mathbb{K}^*_{k} = \argmax_{\mathbb{K}_{k}} (\mathcal{C}(\mathbb{K}_{k})),
    \label{up}
\end{equation}
where $\mathcal{C}(\mathbb{K}_{k})$ is the capacity of the subset of users $\mathbb{K}_{k}$. Note that, throughout this paper, we assume that $k$ is known to the BS whereas in practice, finding the optimal $k$ is a different optimization problem in itself. For instance, in real networks, MU-MIMO user pairing is achieved for 8, 16 users, etc \cite{survey}. Ideally, \eqref{up} requires an exhaustive search of all possible sets $\mathbb{K}_{k}$ to find an optimal solution which is an NP-hard problem. However, this problem can be solved without performing an exhaustive search by exploiting the channel characteristics and the correlation between different users. 

The MU-MIMO technology aims to transmit independent data streams to different users, but only a portion of the transmitted data symbols are useful for each user. A critical issue in this scenario is mitigating inter-user interference (IUI), which involves suppressing information intended for other receivers. This interference highly depends on the extent of overlap among the subspaces spanned by channel matrices of different users. In other words, pairing users with the least amount of overlap would result in an optimal solution.

Consider a single user $k$, whose interference matrix can be given as  $\bm{\tilde{H}}_{k} = [\bm{h}_{1}^T, \hdots, \bm{h}_{k-1}^T,\\ \bm{h}_{k+1}^T, \hdots, \bm{h}_{K}^T]^T$. We define the subspace spanned by the other $K-1$ interfering users as $\mathcal{V}_{k} = Span(\bm{\tilde{H}}_{k})$ and let $\mathcal{V}_{k}^{\perp} = Span(\bm{\tilde{H}}_{k}^{\perp})$ be a subspace orthogonal to $\mathcal{V}_{k}$ which is also the null space of the interference matrix $\bm{\tilde{H}}_{k}$. Now, the matrix $\bm{\tilde{H}}_{k}$ can be projected onto the subspaces $\mathcal{V}_{k}$ and $\mathcal{V}_{k}^{\perp}$ as $\bm{\tilde{H}}_{k} = \bm{\tilde{H}}_{k}^{||} + \bm{\tilde{H}}_{k}^{\perp}$, where $\bm{\tilde{H}}_{k}^{||}$ represents the attenuation caused by correlation among interfering users, and $\bm{\tilde{H}}_{k}^{\perp}$ represents the component that is free of IUI from the rest $K-1$ users. This is known as the zero-forcing direction or the null space projection \cite{overview1}. Finding an optimal solution $\mathbb{K}^*_{k}$ entails a series of null space projections to eventually find a direction that is free of IUI. However, the existence of this direction is not guaranteed as the channels may be highly correlated. Consequently, beamforming or precoding is used to separate the channels by intelligently selecting the optimal set of users, $\mathbb{K}^*_{k}$. Zero Forcing Beamforming is one such precoder that can find the zero-forcing direction.

\subsection{Zero Forcing Beamforming (ZFBF)} 
In Zero Forcing Beamforming (ZFBF), the interference is canceled by choosing beamforming vectors as $\bm{h}_{i}\bm{w}_{j}=0$ for $j\neq i$. If $\mathbb{K}_{k}$ is the set of scheduled users and $\bm{H}(\mathbb{K}_{k})$ and $\bm{W}(\mathbb{K}_{k})$ are the corresponding channel and precoding matrices which are the submatrices of $\bm{H} = [\bm{h}_{1}^T, \hdots, \bm{h}_{K}^T]^T$ and $\bm{W} = [\bm{w}_{1}, \hdots, \bm{w}_{K}]$ respectively. To obtain zero interference, $\bm{W}(\mathbb{K}_{k})$ can be chosen as the pseudo inverse of $\bm{H}(\mathbb{K}_{k})$ as follows\cite{sus},
\begin{equation}
    \bm{W}(\mathbb{K}_{k}) = \bm{H}(\mathbb{K}_{k})^\dag = \bm{H}(\mathbb{K}_{k})^*\big(\bm{H}(\mathbb{K}_{k})\bm{H}(\mathbb{K}_{k})^*\big)^{-1}\nonumber.
    \label{capacity}
\end{equation}
The sum rate achieved by scheduling $k$ users by any beamforming scheme is given by
\begin{equation}
    \mathcal{C}(\mathbb{K}_{k}) = \max_{\bm{w}_{i}, P_{i}}\sum_{i = 1}^{k}\log_{2}\Bigg(1+\frac{P_{i}|\bm{h}_{i}\bm{w}_{i}|^2}{1+\sum_{j=1, j\neq i}^{k}P_{j}|\bm{h}_{i}\bm{w}_{j}|^2}\Bigg),
    \label{sumrate}
\end{equation}
which is subject to $\sum_{i=1}^{k}|\bm{w}_{i}|^2P_{i} \leq P$, where $P$ is the total power and $P_{i}$ is the power allocated to the $i$-th user. $P_{i}$'s can be obtained by using the water-filling algorithm \cite{sus}. However, with ZFBF, the sum rate simplifies to 
\begin{equation}
    \mathcal{C}(\mathbb{K}_{k}) = \max_{P_{i}}\sum_{i = 1}^{k}\log_{2}\big(1+P_{i}\big) \nonumber.
    \label{capacity_zfbf}
\end{equation}
Finally, the maximum achievable sum rate by ZFBF is again given by \eqref{up}. Note that, this still needs an exhaustive search which is computationally expensive. To circumvent this, user grouping was introduced where users with high correlation who cannot be co-scheduled are grouped together. Thereafter, users are scheduled in such a way that, ideally, no two users belong to the same group. This reduces the complexity of finding the optimal solution of \eqref{up} as the search space is reduced. Correlation or overlap between subspaces spanned by users can be quantified by distance metrics like chordal distance, the angle between subspaces, coefficient of correlation, etc \cite{overview}. $k$-means clustering using chordal distance has been a popular approach to achieve user grouping \cite{jsdm1} \cite{prev1}.
\subsection{k-means}
User pairing can also be achieved by using the $k$-means clustering algorithm \cite{ug} \cite{jsdm1}. In $k$-means, observations are clustered into groups such that they belong to the cluster with the nearest mean (center). To find $\mathbb{K}_{k}$ of size $k$, firstly, we need to cluster all $K$ users into $k$ groups. Secondly, we pick one user from each group that is closest to the cluster center. These $k$ users may not form the optimal subset $\mathbb{K}^*_{k}$ but they give statistically good throughput at a low cost of computation. In other words, $k$-means exploits the trade-off between complexity and accuracy to generate a sub-optimal solution much faster (takes polynomial time) than an exhaustive search employing ZFBF (takes exponential time).

As mentioned earlier, we use chordal distance as the similarity measure to group users. For that, observation space consists of the $K$ user covariance dominant eigenspaces, i.e., $\{U_{k}:k = 1, \hdots, K\}$. The chordal distance $d_{C}(i,j)$ between the users $i$ and $j$ is defined as 
\begin{equation}
    d_{C}(i,j) = ||\bm{U}_{i}\bm{U}_{i}^{H} - \bm{U}_{j}\bm{U}_{j}^{H}||_{F}^2 \nonumber,
    \label{chordal}
\end{equation}
which is the distance between the covariance eigenspaces of users $i$ and $j$, where $||\cdot||_{F}$ denotes the Frobenius norm. Using this distance metric, $k$-means is performed and once the grouping is done, picking the $k$ users closest to their respective cluster centers is straightforward and scheduling is achieved.

Although $k$-means scheduling is faster than a brute force exhaustive search on the user space, it still has several disadvantages, namely:
\begin{enumerate}
    \item Sensitivity: The k-means algorithm is sensitive to the initial choice of cluster centers, which can lead to different results for different instances. The most generic $k$-means algorithm uses a random initialization method which may not always find the optimal solution.
    \item Convexity: $k$-means is designed to identify convex clusters, and it may not be suitable for non-convex or irregularly shaped clusters.
    \item Uniformity: $k$-means assumes that all clusters have equal variances and sizes which is usually not true for user distributions.
    \item Convergence: $k$-means may not always converge to a solution, or it may converge to a local minimum instead of the global minimum which also depends on the choice of initial clusters.
\end{enumerate}
There exist several variants of $k$-means clustering, addressing each of these issues \cite{ksurvey}. Nonetheless, it is still an iterative approach that does not scale well for large-scale networks. To address these issues, we propose to use GNNs for user pairing as they provide an efficient framework to model the structure and dynamics of wireless networks.

\begin{figure}[h]
\centering
\includegraphics[width=0.5\linewidth]{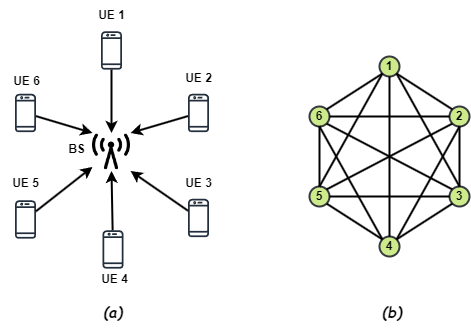}
\captionof{figure}{(a) Illustration of a wireless network with a Base Station (BS) serving 6 Users (UEs). (b) Representation of the wireless network in (a) as a graph.} 
\label{wcg}
\end{figure}

\section{Problem Formulation}
In this section, motivated by the advantages of GNNs in representing graph data, we formulate the user pairing problem using graph theory.
\subsection{Graph} A generic weighted graph can be represented as $G = (V, E, W)$, where $V$ is the set of nodes, $E$ is the set of edges, and $W$ is the set of edge weights. We define $v_{i} \in V$, $e_{ij} = (v_{i}, v_{j}) \in E$ and $w_{i,j} \in W$  to denote a node, an edge between $v_{i}$ and $v_{j}$, and edge weight between $v_{i}$ and $v_{j}$ respectively. The neighborhood of a node $v$ is defined as $N(v) = \{u \in V |(v, u) \in E\}$. A graph may have node attributes \textbf{X}, where \textbf{X} $\in$ \textbf{R}$^{n\times d}$ is a node feature matrix with $x_{v} \in$ \textbf{R}$^{d}$ representing the feature vector of a node $v$ and $n$ being the total number of nodes in $G$. \par
\subsection{Construction of Wireless Communication Graph}
With the framework of graphs introduced, we define the representation of a wireless network as a graph. In a scenario with only one base station (BS) that allocates resources like power control and user association as shown in Fig.{~\ref{wcg}}, the UEs are treated as nodes forming a wireless communication graph (WCG). The $i$-th UE is considered a node, and its feature vector should ideally contain information about CSI and other statistical information. The weight ($w_{ij}$) of the edge between nodes $v_i$ and $v_j$ should include information about the CSI and inter-user interference (IUI). In general, chordal distance \cite{chord} is used as the edge weight for measuring similarity between users in order to capture IUI or the overlap of subspaces spanned by the users.

\subsection{Sum Rate Distance}
In this work, we introduce a novel way to calculate the edge weights of a WCG that facilitates formulating the user pairing problem with GNNs. In this method, the edge weight between $i$-th and $j$-th user of a WCG is defined as the sum rate achieved by the two users $i$ and $j$ when scheduled together which is (from \eqref{sumrate}):
\begin{align}
d_{S}(i,j)  = \max_{\bm{w}_{i}, P_{i}, \bm{w}_{j}, P_{j}}\Bigg(\log_{2}\bigg(1+\frac{P_{i}|\bm{h}_{i}\bm{w}_{i}|^2}{1+P_{j}|\bm{h}_{i}\bm{w}_{j}|^2}\bigg)+ \nonumber \\
 \log_{2}\bigg(1+\frac{P_{j}|\bm{h}_{j}\bm{w}_{j}|^2}{1+P_{i}|\bm{h}_{j}\bm{w}_{i}|^2}\bigg)\Bigg),
 \label{ds}
\end{align}

which is subject to $|\bm{w}_{i}|^2P_{i} + |\bm{w}_{j}|^2P_{j}  \leq P. \delta$ where $\delta = \frac{k-1}{{k \choose 2}}$ and $k$ is the total number of users to be scheduled in the WCG. The value of $\delta$ is selected to maintain the total power available at the BS, as derived in Appendix A. Let's call $d_{S}(i,j)$ as the \textit{sum rate distance}. This is defined in this way in order to find the optimal solution, $\mathbb{K}^*_{k}$ for a ZFBF precoder ($\bm{h}_{i}\bm{w}_{j}=0$ for $j\neq i$). For instance, consider a network with $K$ users and $k$ scheduled users. With this new distance metric, the capacity of this system for any $\mathbb{K}_{k}$ with ZFBF precoder can be given as (derived in Appendix B),
\begin{equation}
    \mathcal{C}(\mathbb{K}_{k}) = \frac{1}{k-1}\sum_{i,j \in \mathbb{K}_{k}}d_{S}(i,j).
    \label{cap_new}
\end{equation}
From \eqref{up} and \eqref{cap_new}, we get
\begin{equation}
    \mathbb{K}^*_{k} = \argmax_{\mathbb{K}_{k}} \Big(\sum_{i,j \in \mathbb{K}_{k}}d_{S}(i,j)\Big).
    \label{new_up}
\end{equation}

\begin{figure}[h]
\centering
\includegraphics[width=0.5\linewidth]{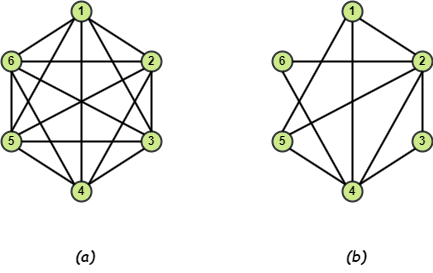}
\captionof{figure}{(a) Wireless communication graph (WCG) of 6 UEs. (b) Sparsified WCG after removing some edges.} 
\label{sparse}
\end{figure}
By using the sum rate distance metric, we established a direct relationship between the total edge weight of the subgraph formed by scheduling users $\mathbb{K}_{k}$, i.e., $\sum_{i,j \in \mathbb{K}_{k}}d_{S}(i,j)$, and the capacity of the system by employing ZFBF precoding, i.e., $\mathcal{C}(\mathbb{K}_{k})$ as shown in \eqref{new_up}. Therefore, capacity can be optimized by choosing a subgraph of size $k$ with the highest total edge weight. However, any WCG is a complete graph, and calculating the total weight of all possible subgraphs of size $k$ is an NP-hard problem \cite{erdos}. Therefore, to reduce the complexity of the problem, we remove some connections from the WCG to make it sparser as shown in Fig.{~\ref{sparse}}. We achieve this by redefining the edges, $E$ in the graph $G = (V, E, W)$ as follows
\begin{equation}
    E(G) = \{e_{i,j}: e_{i,j} \in E, w_{i,j} > \beta \times \Bar{w}\},
    \label{mod}
\end{equation}
where, $\Bar{w}$ is the average edge weight of the WCG and $\beta$ is the sparsifying coefficient. If all the edge weights of the WCG are normalized to 1, then $\beta \in [0,\frac{1}{\Bar{w}})$. To solve the subgraph optimization problem in polynomial time, sparsification of WCG becomes necessary though it reduces the accuracy. Now that we have set up the WCG, we next formulate this optimization problem.

\subsection{k-clique optimization}
For a weighted graph like WCG, we model the user pairing problem as a $k$-clique problem which is a well-known construct in graph theory that involves finding the set of $k$ vertices in a graph with the largest sum of edge weights, where each vertex is connected to every other vertex (an example of $6$-clique is shown in Fig.{~\ref{wcg}}). Therefore, user pairing can be represented by a $k$-clique optimization problem to find a complete sub-graph with $k$ nodes that have the highest sum of edge weights. We define $\Omega_{k}$ as the family of $k$-cliques of the modified WCG as per \eqref{mod}. With this, the user pairing problem can be formally defined as 
\begin{align}
    \mathbb{K}^*_{k} &= \{s: s\in S^*, S^* \subseteq V\}, \quad \mbox{where} \nonumber\\ S^* &= \argmax_{S} \sum_{v_{i},v_{j} \in S} w_{ij}, \quad \mbox{where} \quad S \in \Omega_{k}.
    \label{k_up}
\end{align}
This will optimize the capacity under ZFBF due to the sum rate distance metric introduced in Section IV-C.

\section{User pairing using GNNs}
In the previous section, we formulated the user pairing problem in terms of graph theory as a $k$-clique optimization problem. In this section, we discuss the necessary tools to solve this optimization problem using GNNs.
\subsection{Maximum-clique optimization}
GNNs have shown promise in addressing another variant of this problem called maximum-clique optimization which is to find the largest set of nodes that form a clique in a graph. Finding the maximum clique is proved to be NP-complete \cite{erdos}. GNNs \cite{erdos} solve this by representing the graph as a set of node features and leveraging neural network techniques to learn a mapping from these features to the probability of each node being part of the maximum clique. Since labeling the user data for training GNNs is difficult, we resort to unsupervised models such as Erdős' probabilistic method introduced in \cite{erdos} to solve combinatorial optimization problems using GNNs.

Erdős probabilistic theory is based on the idea of using probabilistic methods to solve problems in combinatorics, number theory, graph theory, and other areas of mathematics. The main idea behind this theory is to use randomness and probability to find unexpected patterns or solutions to difficult problems. It is also used to demonstrate the existence of objects with desired combinatorial properties \cite{erdos,erdos1,erdos2}. To use this method, let's define $\Omega_{max}$ as the family of maximum cliques of the modified WCG as per \eqref{mod}. We then redefine the user pairing problem as follows
\begin{align}
    \mathbb{K}^*_{max} &= \{s: s\in S^*_{max}, S^*_{max} \subseteq V\}, \quad \mbox{where} \nonumber \\ S^*_{max} &= \argmax_{S_{max}} \sum_{v_{i},v_{j} \in S_{max}} w_{ij},\quad \mbox{where} 
 \quad S_{max} \in \Omega_{max}.
    \label{max_up}
\end{align}
\begin{figure}[h]
\centering
\includegraphics[width=0.6\linewidth]{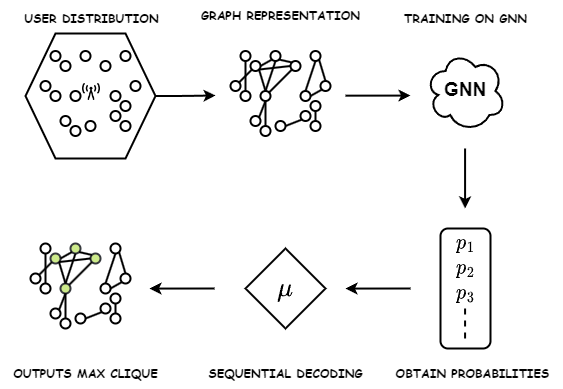}
\captionof{figure}{Erdős goes neural pipeline illustrating training phase of the GNN and sequential decoding phase.} 
\label{pipeline}
\end{figure}
\eqref{max_up} is different from \eqref{k_up} in that, here, we schedule the maximum number of users possible instead of $k$ users as studied earlier. The objective of scheduling the maximum number of users can be achieved by identifying the clique that has the highest number of nodes and maximum total edge weight in a graph. This approach may seem preferable to the $k$-clique optimization since it eliminates the need to determine the value of $k$ by automatically scheduling the maximum number of users. However, maximum-clique optimization presents several challenges in practice, which will be addressed in subsequent sections.

Now that, we have slightly modified the user pairing problem to better suit the techniques adopted in \cite{erdos}, we restate the \textit{"Erdős goes neural"} pipeline here as follows:
\subsection{Erdős goes neural pipeline}
\eqref{max_up} is non-differential, hence, it cannot be directly employed as a loss function to train GNNs. Therefore we resort to the probabilistic method where we train a GNN to output a probability distribution over the nodes determining whether they belong to a maximum clique. Erdős goes Neural \cite{erdos} pipeline comprises three steps after forming the WCG, as depicted in Fig.{~\ref{pipeline}}. 

\begin{enumerate}
    \item Given a WCG, define a GNN $g_{\Theta}$ that outputs a distribution $D = g_{\Theta}(G)$ over the nodes, determining the probability of each node belonging to the maximum-clique solution, where $\Theta$ are the trainable parameters in the network.
    \item Define a loss function as in \eqref{loss} using the probabilistic method and then train the GNN $g_{\Theta}$ to optimize the probability that there exists a valid distribution $D$ or a solution set $S^*_{max} (\in \Omega_{max})$ with a large total edge weight $\sum_{v_{i},v_{j} \in S_{max}} w_{ij}$.
    \item Sequentially obtain an integral solution $S^*_{max}$ from the probability distribution $D$ by sequential decoding as given by \eqref{seq}.
\end{enumerate}
Before training, $D$ is simply initialized to a Bernoulli distribution where the decision of whether $v_{i} \in S_{max}$ is determined by a Bernoulli random variable $x_{i}$ of probability $p_{i}$. 
\subsubsection{Loss Function}
The probabilistic penalty loss as defined in \cite{erdos} for the maximum-clique problem is given as:

\begin{equation}
    \mathcal{L}_{max} = \gamma - (\beta+1)\sum_{(v_{i},v_{j})\in E} w_{ij}p_{i}p_{j} + \frac{\beta}{2} \sum_{v_{i}\neq v_{j}} p_{i}p_{j}
    \label{loss},
\end{equation}
where $\gamma$ and $\beta$ satisfy $\max_{S_{max}}(\sum_{v_{i},v_{j} \in S_{max}} w_{ij}) \leq \gamma \leq \beta$ and $w_{ij} \leq 1$. It should be noted that minimizing this loss function has two effects. Firstly, it automatically maximizes the expectation value of the sum of all edge weights in the graph, represented by $\sum_{(v_{i},v_{j})\in E} w_{ij}p_{i}p_{j}$. Secondly, it minimizes the sum of pairwise products of the probabilities of all the nodes, which is denoted by $\sum_{v_{i}\neq v_{j}} p_{i}p_{j}$. By combining these effects, the loss function reduces the probability of nodes that are not connected by edges and increases the probability of nodes with higher edge weights. This loss function is therefore suitable for maximum clique optimization. Combined with sequential decoding which is defined in the following section, maximum-clique optimization can be solved in polynomial time.

\subsubsection{Sequential Decoding}
To deterministically obtain $S^*_{max}$ with integral solutions, \cite{erdos} proposed to use the following method:
\begin{enumerate}
    \item After obtaining the probability distribution over the nodes, D, arrange the nodes in decreasing order of their probabilities. Sort the indices accordingly and store them in the array $\bm{\phi}$.
    \item In the first iteration, $i=0$, set $S^*_{max} = \O$.
    \item In the $i$-th iteration add the $i$-th sorted node, i.e., $v_{\bm{\phi}(i)}$ to the set $S^*_{max}$ if $i$-th sorted node is connected to all the other $|S^*_{max}|-1$ nodes. This update rule to add $v_{\bm{\phi}(i)}$ to $S^*_{max}$ is given by,
\end{enumerate}
\begin{equation}
    S^*_{max} = S^*_{max} \cup \bigg\{v_{\bm{\phi}(i)} : |E(S^*_{max} \cup \{v_{\bm{\phi}(i)}\} )| = {|S^*_{max}| \choose 2}\bigg\},
    \label{seq}
\end{equation}
conditioned upon $v_{\bm{\phi}(i)} \in V$ and $v_{\bm{\phi}(i)} \cap S^*_{max} = \O$, where $|\cdot|$ is the cardinality of the set. So far, we described the loss function $\mathcal{L}_{max}$ and the procedure of sequential decoding which altogether helps us in obtaining the maximum clique of the given WCG. Theoretically, this can be used to schedule the maximum number of users while maximizing the sum rate. However, maximum-clique optimization schedules the maximum number of users (which we have no control over) that maximizes the sum rate where each user receives only fractional and limited bandwidth which is not desirable to implement. Thus, designing a scheduling algorithm that can schedule $k$ (that we can input to the algorithm) users becomes essential. To achieve that, we modify the sequential decoding scheme to stop adding the nodes to $S^*_{max}$ in \eqref{seq} when the cardinality reaches $k$. This modified scheme is described as follows

\begin{enumerate}
    \item After obtaining the probability distribution over the nodes, D, arrange the nodes in decreasing order of their probabilities. Sort the indices accordingly and store them in the array $\bm{\phi}$.
    \item In the first iteration, $i=0$, set $S^*_{k} = \O$.
    \item In the $i$-th iteration add the $i$-th sorted node, i.e., $v_{\bm{\phi}(i)}$ to the set $S^*_{k}$ if $i$-th sorted node is connected to all the other $|S^*_{i}|-1$ nodes and $|S^*_{k}|< k$. This update rule to add $v_{\bm{\phi}(i)}$ to $S^*_{k}$ is given by,
\end{enumerate}
\begin{equation}
    S^*_{k} = S^*_{k} \cup \bigg\{v_{\bm{\phi}(i)} : |E(S^*_{k} \cup \{v_{\bm{\phi}(i)}\} )| = {|S^*_{k}| \choose 2}, |S^*_{k}|< k\bigg\},
    \label{seq}
\end{equation}
conditioned upon $v_{\bm{\phi}(i)} \in V$ and $v_{\bm{\phi}(i)} \cap S^*_{max} = \O$, where $|\cdot|$ is the cardinality of the set. The proposed approach to obtaining the optimal user pairing  is summarized as follows

\begin{algorithm}[h]
\caption{User Pairing}
\begin{algorithmic}
\REQUIRE $\mathbb{K}^*_{k} = \{s: s\in S^*, S^* \subseteq V\}$ as given by \eqref{k_up}.
\STATE \textit{1) Represent the user distribution as a WCG.}  
\STATE \textit{2) Sparsify the WCG by removing edges as per} \eqref{mod}.
\STATE \textit{3) Define the node feature matrix $\bm{X}$ as an all-ones \\matrix which is a standard initialization.}  
\STATE \textit{4) Train the GNN $g_{\Theta}$ with the loss function given in \eqref{loss}.}  
\STATE \textit{4) Obtain the node probabilities and sort them in decreasing order. Store the sorted indices in the array $\bm{\phi}$.}
\STATE \textit{5) Perform sequential decoding to obtain the optimal $k$-clique subgraph as per \eqref{seq}.} 

\end{algorithmic}
\end{algorithm}

Now that we designed the scheme to identify the optimal $k$-clique from a WCG, we describe the GNN architecture used to deploy this scheme.
\subsection{GNN Architecture}
\begin{figure}[h]
\centering
\includegraphics[width=0.6\linewidth]{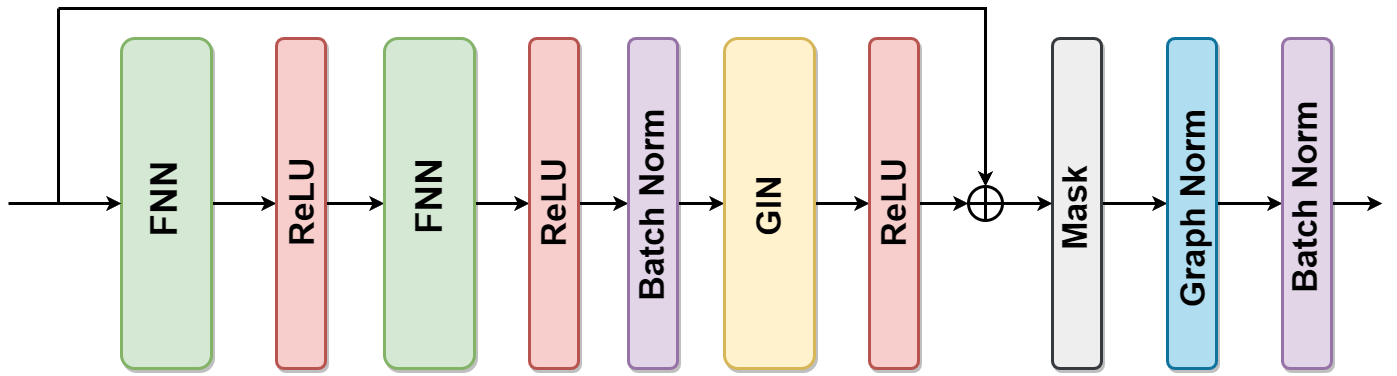}
\captionof{figure}{Fundamental convolution block (FCB) of the proposed GNN} 
\label{block}
\end{figure}

Graph Isomorphism Network (GIN) layers \cite{base} perform the convolution operation for the WCG. Graph size normalization for each convolution was incorporated, as it improved optimization stability. The receptive field was masked to ensure locality at each layer. This means that only 1-hop neighbors were allowed to have nonzero values after one layer of convolution, while 2-hop neighbors were permitted to have nonzero values after two layers, and so on. This series of operations constitute the fundamental convolution block (FCB) of the GNN architecture we used to solve the $k$-clique optimization problem as shown in Fig.{~\ref{block}}. For the one-ring channel model, we used 8 such blocks in series.

After the final GNN layer, a two-layer perceptron was used to generate a single output value for each node. To scale the resulting numbers within the range of $[0, 1]$, a graph-wide min-max normalization was applied. These normalized values, denoted as $p_{1},\hdots, p_{n}$, were then treated as probabilities.

\section{Numerical Results}
With the above-described GNN architecture and problem formulation, we performed a series of simulations on the one-ring channel model to obtain the following results \footnote{Source code of this paper: \url{https://github.com/sharanmourya/UserPairing}}.
\subsection{Sum Rate Performance}
\begin{figure}[h]
\centering
\includegraphics[width=0.6\linewidth]{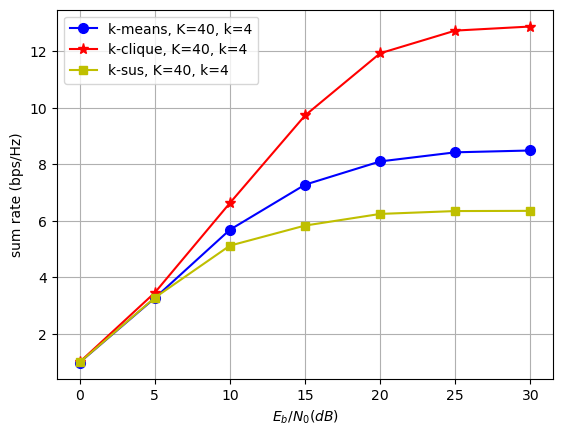}
\captionof{figure}{ Comparison of the proposed $k$-clique method with $k$-means and $k$-SUS (semi-orthogonal user selection). Sum rate vs $E_{b}/N_{0}$ curves for each method are plotted for $k = 4$ users.} 
\label{sr_vs_snr}
\end{figure}

In essence, the user pairing problem can be reduced to maximizing the sum rate for a fixed value of $k$. To evaluate our proposed approach against existing methods, we conducted extensive simulations using a one-ring channel model with $K=40$, $k=4$, and $\beta=1$. The results are plotted in Fig.{~\ref{sr_vs_snr}} to show the sum rate as a function of $E_{b}/N_{0}$.
\begin{figure}[h]
\centering
\includegraphics[width=0.6\linewidth]{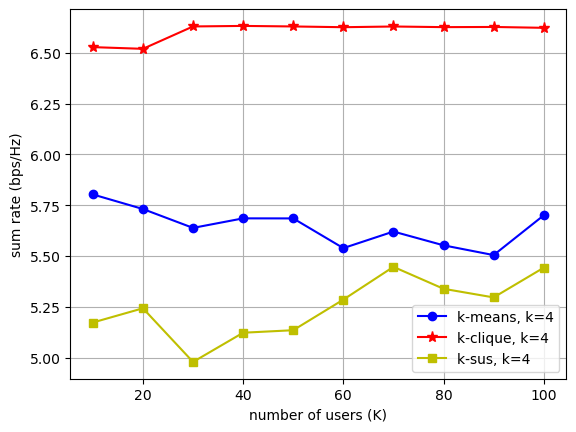}
\captionof{figure}{Comparison of the proposed $k$-clique method with $k$-means and $k$-SUS (semi-orthogonal user selection). Sum rate vs $K$ curves for each method are plotted for $k = 4$ users. } 
\label{snr_vs_K}
\end{figure}
\begin{figure}[h]
\centering
\includegraphics[width=0.6\linewidth]{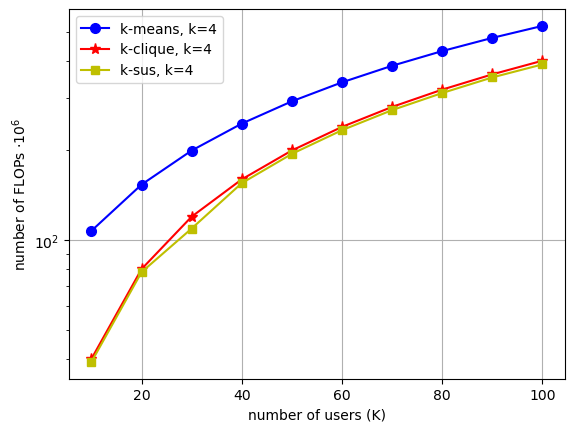}
\captionof{figure}{Comparison of the proposed $k$-clique method with $k$-means and $k$-SUS (semi-orthogonal user selection). The number of FLOPs vs $K$ curves for each method is plotted for $k = 4$ users.} 
\label{flops_vs_K}
\end{figure}

The comparison of our proposed method with $k$-means scheduling based on chordal distance \cite{jsdm1} and semi-orthogonal user pairing (SUS) \cite{sus} is shown in Fig.{~\ref{sr_vs_snr}}. It is evident from the figure that the $k$-clique optimization proposed in this paper outperforms $k$-means and $k$-SUS significantly. At 20 dB SNR, $k$-clique achieved a sum rate of 12 bps/Hz while $k$-means only achieved 8.02 bps/Hz, and $k$-SUS achieved a mediocre 6.14 bps/Hz. The performance of $k$-clique was $49\%$ better than $k$-means and a staggering $95\%$ better than $k$-SUS at 20 dB SNR.

\subsection{Sum Rate Stability}
Besides evaluating the sum rate performance, we also conducted simulations to examine how the sum rate varies with the total number of users in the system ($K$) while keeping the number of scheduled users ($k=4$) constant. Fig.{~\ref{snr_vs_K}} displays the outcomes obtained for a one-ring channel model. The graph reveals that $k$-clique has minimal fluctuations in performance compared to $k$-means or $k$-SUS, as these methods heavily depend on the initial choice of clusters and pivot \cite{sus}, respectively.

\subsection{Complexity}
Comparing the efficiency of various approaches involves complexity as a significant metric. To this end, we determine the theoretical time complexity of the three methods utilized in this research and validate the computations by obtaining simulation outcomes.
\subsubsection{\textbf{k-clique}}
The proposed method consists of several steps, each with a different time complexity:

\begin{enumerate}
    \item Graph Setup: We calculate the sum rate distance between $K$ users, which requires $O(K^2)$ operations since we compute distances for ${K \choose 2} = K*(K-1)/2$ edges.
    \item Edge Reduction: This step involves removing edges with a sum rate distance less than the nominal value ($\beta \times \Bar{w}$ from \eqref{mod}), requiring $O(K)$ operations.
    \item Inference: The graph convolution operation requires $O(K*F^2)$ operations \cite{survey}, where $F$ represents the size of the node embedding vector.
\end{enumerate}
For runtime complexity, we only consider the inference cost, which is $O(K*F^2)$. The linear dependence of the complexity on $K$ is evident from Fig.{~\ref{flops_vs_K}}.
\subsubsection{\textbf{k-means}}
The update equation for $k$-means using chordal distance involves computing the eigenvalue decomposition \cite{jsdm1} of each user to determine the new cluster centers. This process has a maximum complexity of $O(K*M^3)$, where $M$ represents the number of transmit antennas. The linearity of time complexity of $k$-means with respect to $K$ can also be verified from Fig.{~\ref{flops_vs_K}}.

\subsubsection{\textbf{k-SUS}}
The time complexity of semi-orthogonal user pairing is $O(M*K)$, as demonstrated in \cite{sus}. This approach exhibits a linear relationship with respect to the total number of users $K$, which is apparent from Fig.{~\ref{flops_vs_K}}.

\par From Fig.{~\ref{flops_vs_K}}, it is clear that the proposed $k$-clique approach is significantly less complex than $k$-means scheduling in terms of the number of floating point operations per second (FLOPs). At 20 dB SNR, $k$-means consumes a total of 161 MFLOPs whereas $k$-clique consumes only 83 MFLOPs which is $51\%$ of the total FLOPs consumed by $k$-means. In addition, $k$-clique has a very similar complexity compared to $k$-SUS while achieving a $95\%$ improvement in sum rate performance over $k$-SUS at 20 dB SNR.

\subsection{Runtime}
\begin{figure}[h]
\centering
\includegraphics[width=0.6\linewidth]{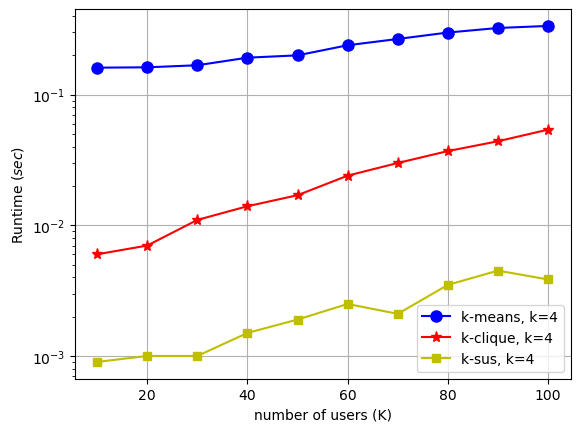}
\captionof{figure}{ Comparison of the proposed $k$-clique method with $k$-means and $k$-SUS (semi-orthogonal user selection). Runtime vs $K$ curves for each method are plotted for $k = 4$ users.} 
\label{runtime_vs_K}
\end{figure}

Runtime measurements for the three approaches are plotted in Fig.{~\ref{runtime_vs_K}}. It can be seen that $k$-SUS is the fastest of the three as it only has matrix multiplication operations. $k$-clique is an order of magnitude faster than the $k$-means scheduling while achieving better overall performance. Although $k$-SUS is the fastest, it lacks in sum rate performance. We conclude that the $k$-clique approach hits the perfect trade-off between complexity and accuracy.

\subsection{Scaling}
GNNs are known for their scalability in processing vast amounts of graph data, even with limited training on smaller graphs. This is mainly due to their ability to use the same parameters across all nodes, making it possible to transfer the trained parameters to a new node without additional training. To demonstrate the scalability of the proposed method, we commence by training the model using the configuration of $K=10$ and $k=4$. Next, we assess the model's performance by varying $K$ from 10 to 100 while maintaining a constant value of $k=4$. 

\begin{figure}[h]
\centering
\includegraphics[width=0.6\linewidth]{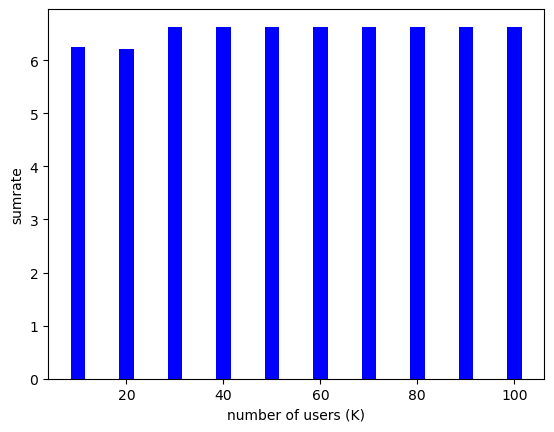}
\captionof{figure}{Scalability of the proposed k-clique method is illustrated by evaluating the GNN for $K$ varying from 10 to 100 when the GNN was explicitly trained for $K=10$. ($k = 4$ is kept constant throughout).} 
\label{scale}
\end{figure}

Fig.{~\ref{scale}} displays the variation of the system's sum rate with respect to $K$ when the model was trained for $K=10$. The plot closely resembles the sum rate graph of $k$-clique in Fig.{~\ref{sr_vs_snr}}, except for a slight reduction in performance overall. This can be attributed to the GNN's exceptional ability to generalize graph data. Although the GNN was not specifically trained for systems with $K\geq 10$, it effectively captured the overall pattern, albeit with slightly lower sum rate values. The decrease in the sum rate for $K\leq 20$ could be due to the network overfitting the data, considering the smaller size of WCG. However, we opted not to reduce the model's trainable parameters or size for smaller networks to maintain consistency across all network sizes. Note that Fig.{~\ref{scale}} only presents the $k$-clique approach as training and inference do not apply to $k$-means and $k$-SUS.

\section{Conclusion}
In this paper, we took an unsupervised learning approach to tackle the problem of user pairing in wireless communication systems by utilizing graph neural networks. Specifically, we employed the Erdős goes neural pipeline, which is well-suited for solving combinatorial optimization problems, to represent the user pairing problem as a $k$-clique optimization. To efficiently solve this optimization problem for a zero-forcing beamforming precoder, we developed a novel distance metric. The resulting model exhibits robustness in terms of performance, complexity, and scalability. We conducted a comparative analysis with semi-orthogonal user scheduling (SUS) and $k$-means scheduling, and our approach demonstrated superior performance while consuming minimal time and resources.

\section*{APPENDIX}
\section*{APPENDIX A.} The sum rate distance metric defined in Section IV-C facilitates formulating user pairing as a $k$-clique problem. The sum rate distance between $i$-th and $j$-th nodes ($d_{S}(i,j)$ as given in \eqref{ds}) is constrained upon $|\bm{w}_{i}|^2P_{i} + |\bm{w}_{j}|^2P_{j}  \leq P. \delta_{i,j}$, where $\delta_{i,j}$ is scalar coefficient for power scaling. Summing all the constraints over the solution set $\mathbb{K}_{k}$ we get,
\begin{align}
    \sum_{i,j \in \mathbb{K}_{k}} \big(|\bm{w}_{i}|^2P_{i} + |\bm{w}_{j}|^2P_{j}\big) &\leq \sum_{i,j \in \mathbb{K}_{k}}P.\delta_{i,j}, \nonumber \\
    \Rightarrow (k-1) \sum_{i \in \mathbb{K}_{k}}\big(|\bm{w}_{i}|^2P_{i}\big) &\leq P\sum_{i,j \in \mathbb{K}_{k}}\delta_{i,j}.
    \label{appa}
\end{align}
Here, we transformed the double summation to a single summation by grouping the terms together, hence the scalar multiplier $(k-1)$. To simplify things lets choose $\delta_{i,j} = \delta$ which results in:
\begin{align}
    (k-1) \sum_{i \in \mathbb{K}_{k}}\big(|\bm{w}_{i}|^2P_{i}\big) &\leq {k \choose 2}P.\delta,  \nonumber \\
    \Rightarrow (k-1).P &\leq {k \choose 2}P.\delta, \nonumber \\  
    \mbox{where} \quad \sum_{i \in \mathbb{K}_{k}}\big(|\bm{w}_{i}|^2P_{i}\big) &\leq P.
\end{align}
By cancelling and rearranging terms we get
\begin{equation}
    \delta \geq \frac{(k-1)}{{k \choose 2}}
\end{equation}
This choice of $\delta$ will make sure that the total power is conserved at the BS.
\section*{APPENDIX B.}
Similar to appendix A, we can derive the capacity of a solution set $\mathbb{K}_{k}$ under ZFBF, in terms of its sum rate distances by adding all the individual edge weights as follows
\begin{align}
    d_{S}(i,j)  &= \max_{P_{i}, P_{j}}\big(log_{2}(1+P_{i})+ 
 log_{2}(1+P_{j})\big), \nonumber \\
    \sum_{i,j \in \mathbb{K}_{k}}d_{S}(i,j)  &= \max_{P_{i}, P_{j}}\sum_{i,j \in \mathbb{K}_{k}}\big(log_{2}(*)+ 
 log_{2}(\#)\big), \nonumber \\
 \mbox{where},\quad (*) &= 1+P_{i}, \quad and \quad (\#) = 1+P_{j}.
\end{align}
Similar to \eqref{appa}, we group similar terms to obtain:
\begin{align}
   \Rightarrow \sum_{i,j \in \mathbb{K}_{k}}d_{S}(i,j)  &= (k-1) \max_{P_{i}}\sum_{i, \in \mathbb{K}_{k}}\big(log_{2}(1+P_{i})\big), \nonumber \\
   \mbox{where}, \quad\mathcal{C}(\mathbb{K}_{k}) &=  \max_{P_{i}}\sum_{i, \in \mathbb{K}_{k}}\big(log_{2}(1+P_{i})\big) \nonumber \\
   \Rightarrow \mathcal{C}(\mathbb{K}_{k}) &= \frac{1}{(k-1)}\sum_{i,j \in \mathbb{K}_{k}}d_{S}(i,j),
\end{align} 
which is the capacity expression given in \eqref{new_up}.

\bibliography{reference}

\begin{thebibliography}{10}

\bibitem{sus}
T.~Yoo and A.~Goldsmith, ``On the optimality of multiantenna broadcast
  scheduling using zero-forcing beamforming,'' {\em IEEE J. Sel. Areas
  Commun.}, vol.~24, no.~3, pp.~528--541, 2006.

\bibitem{ocean}
C.~Luo {\em et~al.}, ``Channel state information prediction for 5g wireless
  communications: A deep learning approach,'' {\em IEEE Trans. Netw. Sci.
  Eng.}, vol.~7, no.~1, pp.~227--236, 2020.

\bibitem{mimo}
T.~J. O'Shea {\em et~al.}, ``Deep learning based mimo communications,'' {\em
  arXiv:1707.07980}, 2017.

\bibitem{stnet}
S.~Mourya {\em et~al.}, ``A spatially separable attention mechanism for massive
  mimo csi feedback,'' {\em IEEE Wireless Commun. Lett.}, pp.~1--1, 2022.

\bibitem{power}
V.~Chien {\em et~al.}, ``Power control in cellular massive mimo with varying
  user activity: A deep learning solution,'' {\em IEEE Trans. Wireless
  Commun.}, vol.~19, no.~9, pp.~5732--5748, 2020.

\bibitem{beam}
T.~Lin and Y.~Zhu, ``Beamforming design for large-scale antenna arrays using
  deep learning,'' {\em IEEE Wireless Commun. Lett.}, vol.~9, no.~1,
  pp.~103--107, 2020.

\bibitem{resource}
M.~Eisen and A.~Ribeiro, ``Optimal wireless resource allocation with random
  edge graph neural networks,'' {\em IEEE Trans. Signal Process.}, vol.~68,
  pp.~2977--2991, 2020.

\bibitem{survey}
Z.~Wu {\em et~al.}, ``A comprehensive survey on graph neural networks,'' {\em
  {IEEE} Trans. Neural Netw. and Learning Syst.}, vol.~32, pp.~4--24, jan 2021.

\bibitem{base}
K.~Xu {\em et~al.}, ``How powerful are graph neural networks?,'' {\em
  arXiv:1810.00826}, 2019.

\bibitem{gnn2}
Y.~Shen {\em et~al.}, ``Graph neural networks for wireless communications: From
  theory to practice,'' {\em IEEE Trans. Wireless Commun.}, pp.~1--1, 2022.

\bibitem{gnn1}
Y.~Shen {\em et~al.}, ``Graph neural networks for scalable radio resource
  management: Architecture design and theoretical analysis,'' {\em IEEE J. Sel.
  Areas Commun.}, vol.~39, no.~1, pp.~101--115, 2021.

\bibitem{overview}
S.~He {\em et~al.}, ``An overview on the application of graph neural networks
  in wireless networks,'' {\em arXiv:2107.03029}, 2021.

\bibitem{gnnw1}
Y.~Shen {\em et~al.}, ``A graph neural network approach for scalable wireless
  power control,'' in {\em 2019 IEEE Globecom Workshops}, pp.~1--6, 2019.

\bibitem{gnnw2}
N.~NaderiAlizadeh {\em et~al.}, ``Adaptive wireless power allocation with graph
  neural networks,'' in {\em IEEE Int. Conf. Acoust. Speech, Signal Process.},
  pp.~5213--5217, 2022.

\bibitem{dec1}
M.~Lee {\em et~al.}, ``Decentralized inference with graph neural networks in
  wireless communication systems,'' {\em IEEE Trans. Mobile Comput.}, vol.~22,
  no.~5, pp.~2582--2598, 2023.

\bibitem{dec2}
Z.~Wang {\em et~al.}, ``Learning decentralized wireless resource allocations
  with graph neural networks,'' {\em IEEE Trans. Signal Process.}, vol.~70,
  pp.~1850--1863, 2022.

\bibitem{gnnw3}
X.~Zhang {\em et~al.}, ``Scalable power control/beamforming in heterogeneous
  wireless networks with graph neural networks,'' in {\em 2021 IEEE Global
  Commun. Conf.}, pp.~01--06, 2021.

\bibitem{prev1}
R.-F. Trifan {\em et~al.}, ``Hybrid mu-mimo precoding based on k-means user
  clustering,'' {\em Algorithms}, vol.~12, no.~7, 2019.

\bibitem{prev2}
S.~Ji {\em et~al.}, ``Deep learning based user grouping for fd-mimo systems
  exploiting statistical channel state information,'' {\em China
  Communications}, vol.~18, no.~7, pp.~183--196, 2021.

\bibitem{jsdm}
A.~Adhikary {\em et~al.}, ``Joint spatial division and multiplexing—the
  large-scale array regime,'' {\em IEEE Trans. Inf. Theory}, vol.~59, no.~10,
  pp.~6441--6463, 2013.

\bibitem{overview1}
Castañeda {\em et~al.}, ``An overview on resource allocation techniques for
  multi-user mimo systems,'' {\em IEEE Commun. Surveys Tuts.}, vol.~19, no.~1,
  pp.~239--284, 2017.

\bibitem{jsdm1}
J.~Nam {\em et~al.}, ``Joint spatial division and multiplexing: Opportunistic
  beamforming, user grouping and simplified downlink scheduling,'' {\em IEEE J.
  Sel. Topics Signal Process.}, vol.~8, no.~5, pp.~876--890, 2014.

\bibitem{ug}
Y.~Xu {\em et~al.}, ``User grouping for massive mimo in fdd systems: New design
  methods and analysis,'' {\em IEEE Access}, vol.~2, pp.~947--959, 2014.

\bibitem{ksurvey}
M.~Ahmed, R.~Seraj, and S.~M.~S. Islam, ``The k-means algorithm: A
  comprehensive survey and performance evaluation,'' {\em Electronics}, vol.~9,
  no.~8, 2020.

\bibitem{chord}
K.~Ko and J.~Lee, ``Multiuser mimo user selection based on chordal distance,''
  {\em IEEE Trans. Commun.}, vol.~60, no.~3, pp.~649--654, 2012.

\bibitem{erdos}
N.~Karalias and A.~Loukas, ``Erdos goes neural: an unsupervised learning
  framework for combinatorial optimization on graphs,'' {\em arXiv:2006.10643},
  2021.

\bibitem{erdos1}
N.~Alon and J.~H. Spencer, ``The probabilistic method,'' {\em John Wiley \&
  Sons}, 2004.

\bibitem{erdos2}
P.~Erdös, ``Graph theory and probability,'' {\em Canadian Journal of
  Mathematics}, vol.~11, p.~34–38, 1959.

\end{thebibliography}
\bibliographystyle{ieeetr}
\end{document}